\documentclass[%
 reprint,
showpacs,
 amsmath,amssymb,
 aps,
 pra,
longbibliography,
floatfix
]{revtex4-1}

\usepackage[utf8]{inputenc}	
\usepackage[T1]{fontenc}	
\usepackage[]{graphicx}		
\usepackage{color}
\usepackage[english]{babel}
\usepackage{hyperref}
\usepackage{url}
\usepackage{todonotes}
\usepackage{lipsum}
\usepackage{bm}

\usepackage{microtype}
\usepackage{braket}

\usepackage[position=top,caption=false]{subfig}

\let\originaleqref\eqref
\renewcommand{\eqref}{Eq.~\originaleqref}

\newcommand{\fref}[1]{\figurename~\ref{#1}}

\renewcommand{\vec}[1]{\bm{#1}}

\newcommand{\Trs}[1]{\mathrm{Tr}(#1)}
\newcommand{\Cvector}[6]
{
\begin{pmatrix}
    #1\\#2\\#3\\#4\\#5\\#6
\end{pmatrix}
}

\newcommand{\sx}[0]{\hat{\sigma}_\mathrm{x}}
\newcommand{\sy}[0]{\hat{\sigma}_\mathrm{y}}

\newcommand{\sm}[0]{\hat{\sigma}_\mathrm{-}}

\newcommand{\Qe}{Q_{\mathrm{e}}}
\newcommand{\rzo}{\rho_{nm}}

\newcommand{\cjz}{\hat{c}_{j,n}}
\newcommand{\cjo}{\hat{c}_{j,m}}
\renewcommand{\H}{\hat{H}}

\newcommand{\psiSEt}[1]{\ket{\psi_{#1}^{\text{SE}}(t)}}
\newcommand{\psiSE}{\ket{\psi_m^{\text{SE}}}}
\newcommand{\cjm}{\hat{c}_{j,m}}

\newcommand{\cjAS}{\hat{c}_{j}^{\text{AS}}}
\newcommand{\pone}{\ket{\psi_1}}
\newcommand{\ptwo}{\ket{\psi_2}}
\newcommand{\pthree}{\ket{\psi_3}}

\newcommand{\verteq}{\rotatebox{90}{$\,=$}}
\newcommand{\equalto}[2]{\underset{\scriptstyle\overset{\mkern4mu\verteq}{#2}}{#1}}

\graphicspath{
{./figures/}
}

\begin{document}
\title{Multi-state and multi-hypothesis discrimination with open quantum systems}
\author{Alexander Holm Kiilerich}
\email{kiilerich@phys.au.dk}
\author{Klaus Mølmer}
\email{moelmer@phys.au.dk}
\affiliation{Department of Physics and Astronomy, Aarhus University, Ny Munkegade 120, 8000 Aarhus C, Denmark}
\date{\today}

\bigskip

\begin{abstract}
We show how an upper bound for the ability to discriminate any number $N$ of candidates for the Hamiltonian governing the evolution of an open quantum system may be calculated by numerically efficient means. 
Our method applies an effective master equation analysis to evaluate the pairwise overlaps between candidate full states of the system and its environment pertaining to the Hamiltonians. These overlaps are then used to construct an $N$-dimensional representation of the states.
The optimal positive-operator valued measure (POVM) and the corresponding probability of assigning a false hypothesis may subsequently be evaluated by phrasing optimal discrimination of multiple non-orthogonal quantum states as a semi-definite programming problem.
We investigate the structure of the optimal POVM and we provide three realistic examples of hypothesis testing with open quantum systems. 
\end{abstract}

\maketitle
\noindent


\section{Introduction}
\begin{figure}
\includegraphics[trim=0 0 0 0,width=0.9\columnwidth]{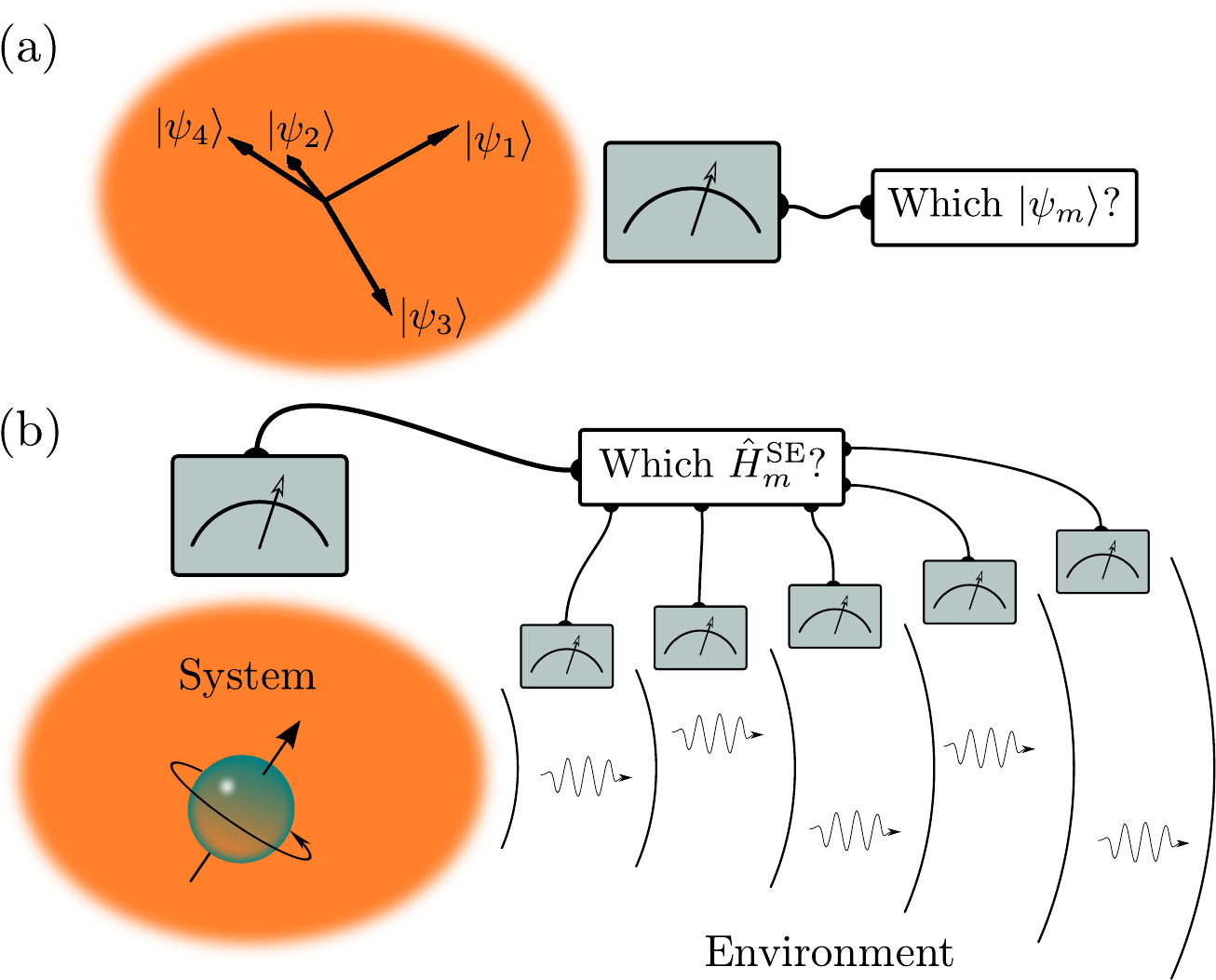}
\caption{
(a) In state discrimination, a measurement is performed to distinguish between a set of $N$ candidate states $\{\psi_m\}_{m=1}^{N}$.
(b) In hypothesis discrimination with an open quantum probe, a combined measurement on the system and its environment is performed to determine the true candidate from a set of $N$ possible Hamiltonians $\{\hat{H}_m^{\text{SE}}\}_{m=1}^{N}$ of the system or the combined system and environment.
}
\label{fig:setup}
\end{figure}

Quantum metrology is concerned with the discrimination of quantum states, \fref{fig:setup}(a), often with the purpose of distinguishing between different physical parameters governing the preparation or evolution of a quantum system \cite{PhysRevLett.96.010401}. In Hamiltonian parameter estimation a continuum of candidate parameter values are filtered by their different action on the state of a quantum probe while hypothesis testing considers scenarios with a discrete set of hypotheses $m = 1,2,\dots, N$, for the Hamiltonian acting on the probe. In both cases, our ability to determine the true candidate hypothesis or physical parameter is ultimately limited by our ability to discern the corresponding signals obtained by measurements on the probe system.

In the present work we are interested in experiments with an open probe system whose interaction with a broadband environment validates the Born-Markov approximation. If the environment is left unmonitored, the interaction leads to decoherence of the system and to a loss of distinguishability while a combined measurement on the probe system and its environment may yield much more information about the physical parameters governing the dynamics, \fref{fig:setup}(b). Assuming that such a measurement is implementable, the ultimate precision is concerned not with the discrimination of reduced density matrix candidates $\rho_m$ for the small system but rather with the discrimination of the, possibly, pure quantum states of the system and its environment $\ket{\psi_m^{\text{SE}}}$.

In this article we show how discrimination between an arbitrary number of (non-orthogonal) states of a quantum system may be employed to determine an upper bound for our ability to distinguish among different hypotheses concerning the evolution of a Markovian open quantum system.
The full states of a system and a Markovian environment occupy in general a very large Hilbert space and the candidate states of the combined system and their overlaps $\braket{\psi_n^{\text{SE}}|\psi_m^{\text{SE}}}$ are at a first glance intractable. Yet, it was shown in Ref.~\cite{PhysRevLett.114.040401}, how $\braket{\psi_n^{\text{SE}}|\psi_m^{\text{SE}}}$ can be calculated efficiently by propagating a so-called two-sided master equation for an effective density matrix which lives in the much smaller Hilbert space of the system alone.

For two candidate states $\pone$ and $\ptwo$ prepared with prior probabilities $P_1$ and $P_2$, Helstrom derived in 1969 a general expression for the minimum error probability in discriminating them by a single measurement \cite{Helstrom1969},
\begin{align}\label{eq:QeMin}
\Qe^{(\text{Helstrom})} = \frac{1}{2}\left(1-\sqrt{1-4P_1P_2|\bra{\psi_1}\psi_2\rangle|^2}\right).
\end{align}
Recent works, e.g.,  \cite{PhysRevA.87.062302,PhysRevA.95.042307}, have made progress towards deriving a general framework for cases with multiple hypotheses, but no closed form expression has been found except in cases where the candidate density operators commute \cite{helstrom1970quantum}. As pointed out by Helstrom \cite{Helstrom1969} it is, however, clear that even for multiple hypotheses, the error probability $\Qe$ depends only on the pairwise overlaps between the candidate states and their prior probabilities.

Our presentation is structured as follows.
In Section~\ref{sec:2} we outline how the error probability in discriminating $N$ arbitrary quantum states can be phrased as a semi-definite-programming problem for which numerically efficient algorithms exist. We then apply the example of discriminating three states of a two-level system to probe the structure of the optimal measurement.
In Section~\ref{sec:3} we re-derive the main results of Ref.~\cite{PhysRevLett.114.040401} for evaluating $\braket{\psi_n^{\text{SE}}|\psi_m^{\text{SE}}}$, and we show how the pairwise state overlaps among $N$ candidate states can be applied to embed these states in a reduced Hilbert space of dimension $N$. 
Distinguishing $N$ hypotheses for the evolution of the open quantum system is equivalent to a multi-state discrimination problem on this Hilbert space. In Section~\ref{sec:4} we illustrate our theory by presenting three examples:
i) Discriminating four candidates for the relative phase of a Rabi drive on an open two-level system,
ii) Discriminating whether a low Q cavity is coupled to $1, 2, 3$ or $4$ atoms, and
iii) Using a sensor ion to determine the position of a nearby qubit ion in a doped crystal lattice structure.
Finally, in Section~\ref{sec:5} we conclude and provide an outlook.

\section{Optimal state discrimination} 
\label{sec:2}
One may specify different goals and hence measures of the quality of a state discrimination process, depending on the number of copies of the quantum system available \cite{davies1978information} and depending on the cost and reward for making wrong and correct estimates \cite{0305-4470-31-34-013}. In the limit where measurements on asymptotically many copies $M$ of the quantum probe system are available, the probability of making an erroneous assignment decreases exponentially with $M$. The exponent obeys the Quantum Chernoff bound \cite{PhysRevLett.98.160501} which was recently generalized to cases with multiple candidate states \cite{li2016discriminating}.

In the present study we are interested in the information obtainable by performing a measurement on a single quantum system. We assume that the system is prepared with probability $P_m$ in one of $N$ different, mixed quantum states $\{\rho_m\}_{m=1}^N$, and that we can perform measurements on the system with outcomes that we combine into our assignment $\lambda = 1,2,\dots,N$ of the most likely state.
We quantify a given measurement strategy by the \textit{error probability} of assigning a false state (hypothesis) based on the outcome $\lambda$ of a measurement performed on the system,
\begin{align}\label{eq:Qe1}
\Qe = \sum_{m=1}^N P_m \mathop{\sum_{n=1}}_{n\neq m}^N P(\lambda=n|\rho_m).
\end{align}
To be able to assign $N$ possible states, the measurement must have $N$ possible outcomes, so for a Hilbert space of dimension $d < N $, we have recourse to generalized (non-projective) measurements with fundamentally ambiguous outcomes. Such measurements are defined by a positive-operator valued measure (POVM) with effects $\{\hat{E}_m\}_{m=1}^N$ which are positive semi-definite ($\hat{E}_m\geq 0$) and sum to identity $\sum \hat{E}_m = \mathbb{I}$, \cite{QMC}.
The probability to obtain an outcome $n$ if the \textit{true} state is $\rho_m$ is then $P(\lambda=n|\rho_m) = \Trs{\hat{E}_n\rho_m}$, so by applying $\sum \hat{E}_m = \mathbb{I}$ and $\sum P_m = 1$ we may rewrite \eqref{eq:Qe1},
\begin{align}
\Qe(\{\hat{E}_m\}_{m=1}^{N}) = 1-\sum_{m=1}^N P_m\Trs{\hat{E}_m\rho_m}.
\end{align}

The task of obtaining the optimal POVM, which minimizes the error probability for a given set of candidate states $\{\rho_m\}_{m=1}^{N}$ with (prior) probabilities $\{P_m\}_{m=1}^{N}$  defines a semi-definite programming problem \cite{vandenberghe1996semidefinite}:
\begin{align}\label{eq:SDP}
\begin{split}
&\text{minimize}\quad \Qe(\{\hat{E}_m\}_{m=1}^{N})
\\
&\text{subject to}\quad \hat{E}_m\geq 0 \quad \forall m \in \{1,\dots,N\}
\\
&\text{and} \quad \sum_{m=1}^N \hat{E}_m = \mathbb{I}.
\end{split}
\end{align}
Reference~\cite{PhysRevA.87.062302} provides an analytic solution for this problem in the case of discriminating three mixed qubit states ($N=3$, $d=2$), but a solution for the general problem of $N$ states in arbitrary Hilbert space dimension has yet to be derived.

A recent study shows how any N-outcome measurement can be decomposed into sequences of nested two-outcome measurements which allows straightforward numerical optimization \cite{PhysRevA.95.042307}.
While such an approach provides insight into the structure of the POVM elements,  we note that since semi-definite programming represents a convex optimization task, the solution can also be directly obtained by numerically efficient algorithms such as Interior Point Methods for small dimensions or first order Conic Optimization which scales to much larger problems at the price of lower precision \cite{vandenberghe1996semidefinite}.
In the present study we apply the CVX package 
for specifying and solving convex programs in Matlab \cite{cvx,gb08}.

\subsection{Example: Three states of a two-level system}
\begin{figure*}\label{fig:states}
\includegraphics[trim=0 0 0 0,width=2\columnwidth]{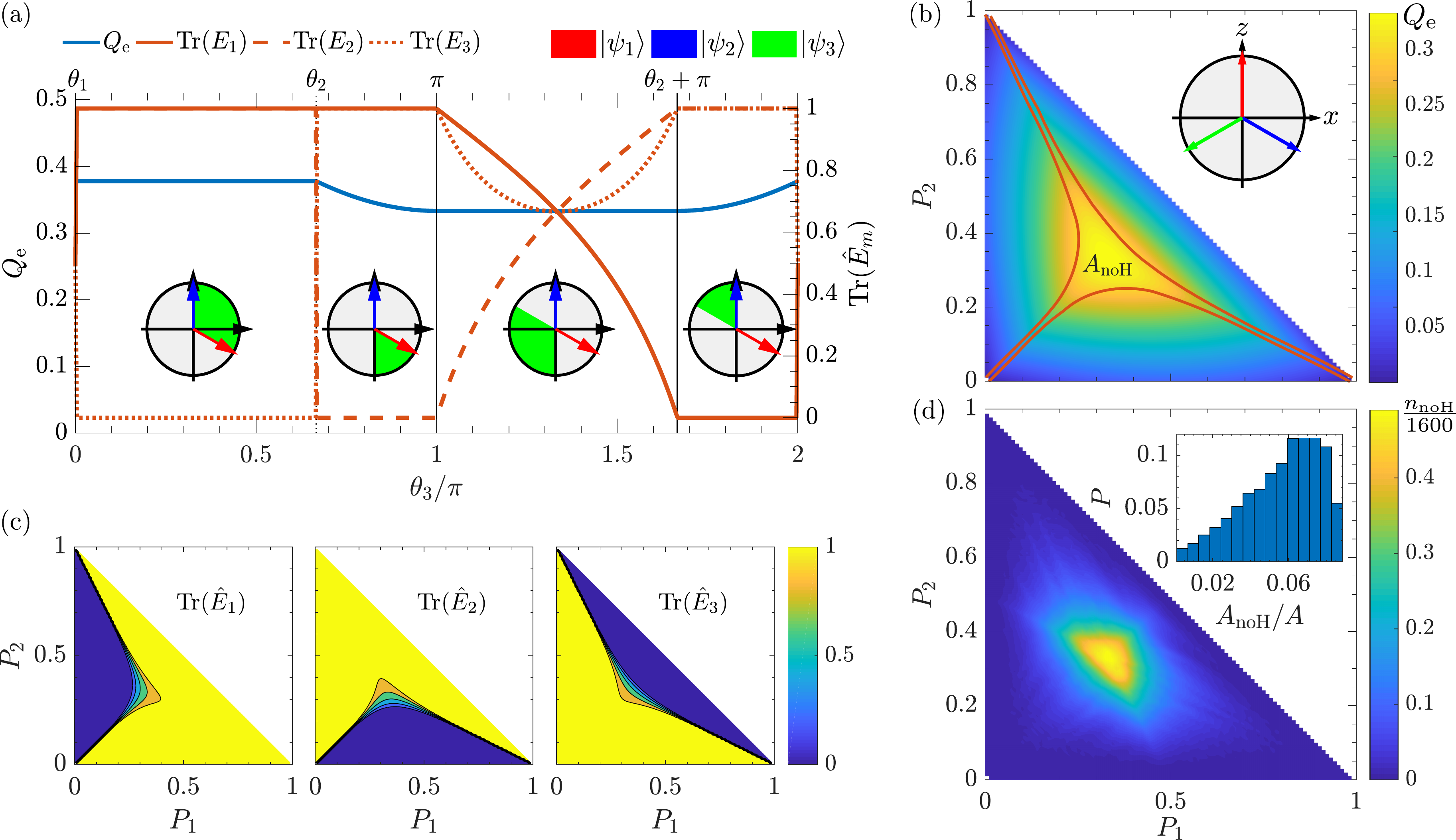}
\caption{
Three candidate two-level states $\pone$, $\ptwo$ and $\pthree$ are distinguished by an optimal measurement. In (a), (b) and (c) the states are restricted to the $xz$-plane of the Bloch sphere.
(a) The left axis (blue line) shows the the error probability and the right axis (red lines) the traces of the POVM effect operators $\hat{E}_m$ as the Bloch vector angle $\theta_3$ of $\ket{\psi_3}$ is varied, while the other candidate states are fixed (see insert figures).
(b) The three states are fixed in a symmetric configuration (see main text and inset), while the prior probabilities $P_1$ and $P_2$ with $P_3 = 1-P_1-P_2$ are scanned from $0$ to $1$. The color plot shows $\Qe$ and the red contours define the boundaries where one of the states can be ruled out and it is optimal to distinguish only between two of the candidate states (the third POVM has $\Trs{\hat{E}_m}< 10^{-4}$, see (c)). Within the red enclosure ($A_{\text{noH}}$) all three candidates must be taken into account. 
(d) Three random pure states on the full Bloch sphere are generated 1600 times. The color plot shows the sampled probability $n_{\mathrm{noH}}/1600$ that a random three-state combination has \textit{all} $\Trs{\hat{E}_m}\geq 0$ as a function of their prior probability weights, and the histogram depicts the distribution of the corresponding areas $A_{\mathrm{noH}}$ where all three states must be distinguished, divided by the total area $A=1/2$.
}
\label{fig:threeStates}
\end{figure*}
As a simple example we show in \fref{fig:threeStates}(a) the error probability $\Qe$ for the discrimination of three pure candidate states $\{\ket{\psi_m}\}_{m=1}^3$ of a two-level system.
We restrict the states to Bloch
vectors $(\sin(\theta_1),0,\cos\theta_1)$, $(\sin\theta_2,0,\cos\theta_2)$ and $(\sin\theta_3,0,\cos\theta_3)$ in the the $xz$-plane of the Bloch sphere. The three candidates are assumed equally probable ($P_m = 1/3$) and we let $\theta_1=0$ and $\theta_2= 2\pi/3$ be fixed while $\theta_3$ is varied from $0$ to $2\pi$.
The plot shows also the traces of the effect operators $\hat{E}_m$ associated with each of the three states. These quantify the importance of being able to obtain an outcome consistent with the candidate $\rho_m$ and, interestingly, we observe that it is often optimal to initially rule out one state and only try to distinguish the two remaining candidates.
In particular, whenever the three states $\ket{\psi_a}$ lie within a semicircle in the Bloch sphere one should disregard one of the hypotheses and obtain the minimum error probability from the Helstrom bound \eqref{eq:QeMin}, while when $\theta_3\in [\pi,\theta_2+\pi]$ all $\Trs{\hat{E}_m}$ are non-zero.

In \fref{fig:threeStates}(b)~and~(c) the tree candidate states are fixed in the $xz$-plane with $\theta_m=2(m-1)\pi/3$ while their prior probabilities are varied.
\fref{fig:threeStates}(b) shows that a lower error $\Qe$ may be achieved when one hypothesis is a priori more likely. In (c) we show the trace of the POVM effect operators. Apart from a relatively small vicinity around $P_1\simeq P_2 \simeq P_3$, the traces are either zero or unity, implying that in most instances one should perform a measurement to discriminate only between the two a priori most likely states while the third is disregarded. Between the edges and the red boundaries in (b), the Helstrom bound \eqref{eq:QeMin} applies and the full numerical solution is only needed in the central area $A_{\mathrm{noH}}$ where a small improvement ($\lesssim 4\%$) over the two-state Helstrom bound is obtained by use of three POVMs.

In \fref{fig:threeStates}(d), three random pure states on the full Bloch sphere are generated 1600 times. For a given set of prior probabilities, the fraction of the samples $n_{\mathrm{noH}}(P_1,P_2)/1600$ with \textit{all} $\Trs{\hat{E}_m}\geq 0$ is shown in the color plot.
The histogram depicts the distribution of the sizes of the corresponding areas $A_{\mathrm{noH}}$ in the space of probabilities relative to the total area $A$.
The area has its largest value ($A_{\mathrm{noH}}/A\simeq 9\%$) for the symmetric combination of states studied in (b) which minimizes the pairwise state overlaps, and the subset of possible prior probabilities where all three states must be taken into account is in general only a modest part of the full set.
The findings presented in relation to \fref{fig:threeStates} confirm the analytic results of Ref.~\cite{PhysRevA.87.062302}.

While we focused our discussion on pure states, we note that similar effects appear if the candidate states are mixed. The purity $\Trs{\rho^2_m}$ then plays a role qualitatively similar to a reduced prior probability.

\section{Hypothesis testing with open quantum systems}
\label{sec:3}
Hypothesis testing is the task of discriminating the evolution of a probe system subject to one of a discrete set of candidate Hamiltonians. Generally, the probe system may be coupled to an environment and the hypotheses concern the total Hamiltonian of the system and its environment.
As argued in \cite{PhysRevLett.112.170401,PhysRevLett.114.040401,PhysRevA.83.062324}, the Markovian nature of the system environment interaction implies that the discernibility of the (unmeasured) quantum states of the system and environment provides a theoretical upper bound for our practical ability to distinguish the different hypotheses by, e.g., continuous monitoring of the environment degrees of freedom as in photon counting or homodyne detection \cite{inprep,PhysRevA.95.022306}.

Each individual hypothesis $m$ leads to a particular unmeasured quantum state $\ket{\psi^{\text{SE}}_m}$ of the system and the environment at a given time $t$ and hypothesis testing is thus equivalent to the problem of discriminating the states $\{\ket{\psi^{\text{SE}}_m(t)}\}_{m=1}^N$, which are in
most cases intractable.
Nevertheless, following the idea of Ref.~\cite{PhysRevLett.114.040401}, we show in this section that in situations where the Born-Markov approximation applies to the system-environment interaction, the overlaps between any two candidate states can be evaluated by solving a two-sided master equation for an effective density operator on the small system Hilbert space alone.
We subsequently show how the overlaps between all pairs of states can be used to construct a low dimensional representation of the problem to which the technique (\ref{eq:SDP}) of Section~\ref{sec:2} applies.

\subsection{A two-sided master equation for the state overlaps}
The distinct hypotheses can be formally mapped to the states $\ket{m}$ of an $N$-level ancillary system such that the evolution of the system and environment is conditioned on the state of the ancilla via the Hamiltonian.
\begin{align}\label{eq:HASE}
\H_{\text{ASE}} = \sum_{m=1}^N \ket{m}\bra{m}\otimes \hat{H}^{\text{SE}}_m.
\end{align}
While the ancilla, system and environment is initially prepared in a separable pure state
$
\frac{1}{\sqrt{N}}\sum_{m=1}^N \ket{m} \otimes \ket{\psi^{\text{SE}}(t=0)},
$
evolution under the Hamiltonian (\ref{eq:HASE}) yields, after a time $t$, an entangled state,
\begin{align}\label{eq:pureFull}
\ket{\psi(t)} = \frac{1}{\sqrt{N}}\sum_{m=1}^N \ket{m}\otimes \psiSEt{m}.
\end{align}
In the Born-Markov approximation, the environmental degrees of freedom can be traced out to yield an effective density matrix $\rho^{(\text{AS})}(t)$ for the (mixed) state of the ancilla and system which takes the form of a block matrix,
\begin{align}
\rho^{\text{AS}}(t) = \frac{1}{N}\sum_{n,m=1}^N \rho_{nm}(t)\ket{n}\bra{m},
\end{align}
where the elements $\rho_{nm}(t)$ of the matrix representation of $\rho^{\text{AS}}(t)$ in the ancilla basis act on the system Hilbert space and are at time $t=0$ identical and given by the initial state of the system.

As seen from \eqref{eq:pureFull}, the overlap between the $n$ and $m$ system-environment candidate states can be obtained as the expectation value of $\ket{n}\bra{m}$,
\begin{align}\label{eq:overlap}
\begin{split}
\braket{\psi^{\text{SE}}_n|\psi^{\text{SE}}_m} &= N\braket{\psi(t)\ket{n}\bra{m}\psi(t)}
\\
&= N\mathrm{Tr}_{\text{AS}}\left[\ket{n}\bra{m}\rho^{\text{AS}}(t)\right]
\\
&=\mathrm{Tr}_{\text{S}}\left[\rho_{mn}(t)\right].
\end{split}
\end{align}
The unitary part of the ancilla-system evolution is governed by a Hamiltonian
\begin{align}
\H_{\text{AS}}  = \sum_{m=1}^N \ket{m}\bra{m}\otimes \hat{H}_m,
\end{align}
where the $\H_m$ are candidates for the part of the system-environment Hamiltonian acting on the system alone.
The interaction with the environment yields an effective set of candidate relaxation operators $\left\{\cjm\right\}_{j=1}^{J}$ for each hypothesis,
and the corresponding evolution of the ancilla-system state adheres to a Lindblad
master equation ($\hbar=1$), $\dot{\rho}^{\text{AS}}= -i[\H_{\text{AS}},\rho^{\text{AS}}]+\sum_j [\cjAS \rho^{\text{AS}} (\cjAS)^\dagger-\frac{1}{2}[(\cjAS)^\dagger \cjAS\rho^{\text{AS}}+\rho^{\text{AS}}(\cjAS)^\dagger \cjAS)]$, where
\begin{align}
\cjAS = \sum_{m=1}^N \ket{m}\bra{m}\otimes \cjm.
\end{align}
It follows that $\rho_{nm}(t)$ solves a \textit{two-sided} master equation
\begin{align}\label{eq:2sided}
\begin{split}
&\frac{d\rzo}{dt} = -i\left(\H_n\rzo-\rzo \H_m\right)
\\ &+
\sum_j\left[\cjz\rzo\cjo^\dagger-\frac{1}{2}\left(\cjz^\dagger\cjz\rzo+\rzo\cjo^\dagger\cjo\right)\right].
\end{split}
\end{align}
The solution of this equation yields by \eqref{eq:overlap} the temporal dynamics of the overlap between any pair of the \textit{full} states of the system and environment pertaining to the different hypotheses for the effective system Hamiltonians $\H_m$ and relaxation operators $\left\{\cjm\right\}_{j=1}^{J}$.
Hence, 
the numerically intractable problem of evolving the full states in the very large system and environment Hilbert space, is reduced to a much simpler task of evolving $(N^2-N)/2$ matrices with the dimension of the smaller probe system.

\subsection{Low dimensional representation of states of the system and its environment}
Although the full state of a system and its environment lives in a formally infinite dimensional Hilbert space, the discrete nature of the hypothesis testing problem implies that at any time we have at most $N$ different possible states to distinguish. These span a (time-dependent) subspace of dimension $N$ which is sufficient to fully characterize the discrimination problem. To apply the semi-definite programming methods of Section~\ref{sec:2}, let us define an orthogonal basis $\{\ket{\phi}_n\}_{n=1}^{N}$ for this subspace such that each candidate state can be expressed as a linear combination,
\begin{align*}
\psiSE= \sum_{n = 1}^m C_n^{(m)} \ket{\phi_n},
\end{align*}
where the $C_n^{(m)}$ are complex expansion coefficients.

We shall now outline, how one may in general define a basis and obtain the $C_n^{(m)}$:
Let the first basis state be the first candidate state
$\ket{\phi_1} = \ket{\psi_1}$, i.e. $C_n^{(1)} = \delta_{n1}$.
The second state is then used to define the second basis state, $\text{span}(\ket{\psi_1^{\text{SE}}},\ket{\psi_2^{\text{SE}}}) = \text{span}(\ket{\phi_1},\ket{\phi_2})$, such that $C_n^{(2)} = 0$ for $n>2$. Since any of the basis states  may be multiplied by an arbitrary complex phase factor, we may further use the convention that $C_2^{(2)}$ is positive which together with the overlap $\braket{\psi_1^{\text{SE}}|\psi_2^{\text{SE}}}$ and the normalization criterion completely determines $\ket{\phi_2}$. Similarly the third candidate defines the third basis state and we set $C_3^{(3)}\in \mathbb{R}_{\geq0}$.
By continuing in this manner, we represent the states of the system and the environment as a sequence of $N$ dimensional vectors
\begin{align}
\begin{split}
\equalto{\quad\vec{C}^{(1)}}{
\Cvector{1}{0}{:}{:}{\vdots}{0}
}
\quad
\equalto{\quad\vec{C}^{(2)}}{
\Cvector{C_{1}^{(2)}}{C_{2}^{(2)}}{0}{:
}{:}{0}
}
\quad
\equalto{\quad\vec{C}^{(3)}}
 {\Cvector{C_{1}^{(3)}}{C_{2}^{(3)}}{C_{3}^{(3)}}{0}{:}{0}
}
\dots
\equalto{\hspace{3pt}\quad\vec{C}^{(N)}}{
\Cvector{C_{1}^{(N)}}{C_{2}^{(N)}}{:}{:}{:}{C_{N}^{(N)}}
}
\end{split}
\end{align}
where all  state amplitudes are given by a recursive procedure:
\begin{align}
C_n^{(m)} &= \frac{1}{C_n^{(n)}}\left(\braket{\psi_n|\psi_m}-\sum_{k=1}^{n-1}C_k^{(n)*}C_k^{(m)}\right)
\end{align}
for $1\leq n\leq m-1$ and
\begin{align}
C_m^{(m)} &= \sqrt{1-\sum_{k=1}^{m-1}|C_k^{(m)}|^2}.
\end{align}

It may in a given hypothesis testing scenario occur that two or several candidate states become identical. The number of POVM elements is then reduced and for some outcomes of our protocol we have recourse to select randomly among the corresponding hypotheses according to their prior probabilities.

\section{Examples}
\label{sec:4}
The ideas and methods presented in Sections~\ref{sec:2}~and~\ref{sec:3} allow us to evaluate the minimum error probability in the assignment of one of \textit{any} number of distinct hypothesis for the evolution of an open quantum system as a function of the duration $t$ of an experiment.
Here we provide three examples which illustrate different aspects of our theory and its application.

\subsection{Phase of a Rabi drive}
\label{sec:Rabi}
\begin{figure}
\includegraphics[trim=0 0 0 0,width=0.9\columnwidth]{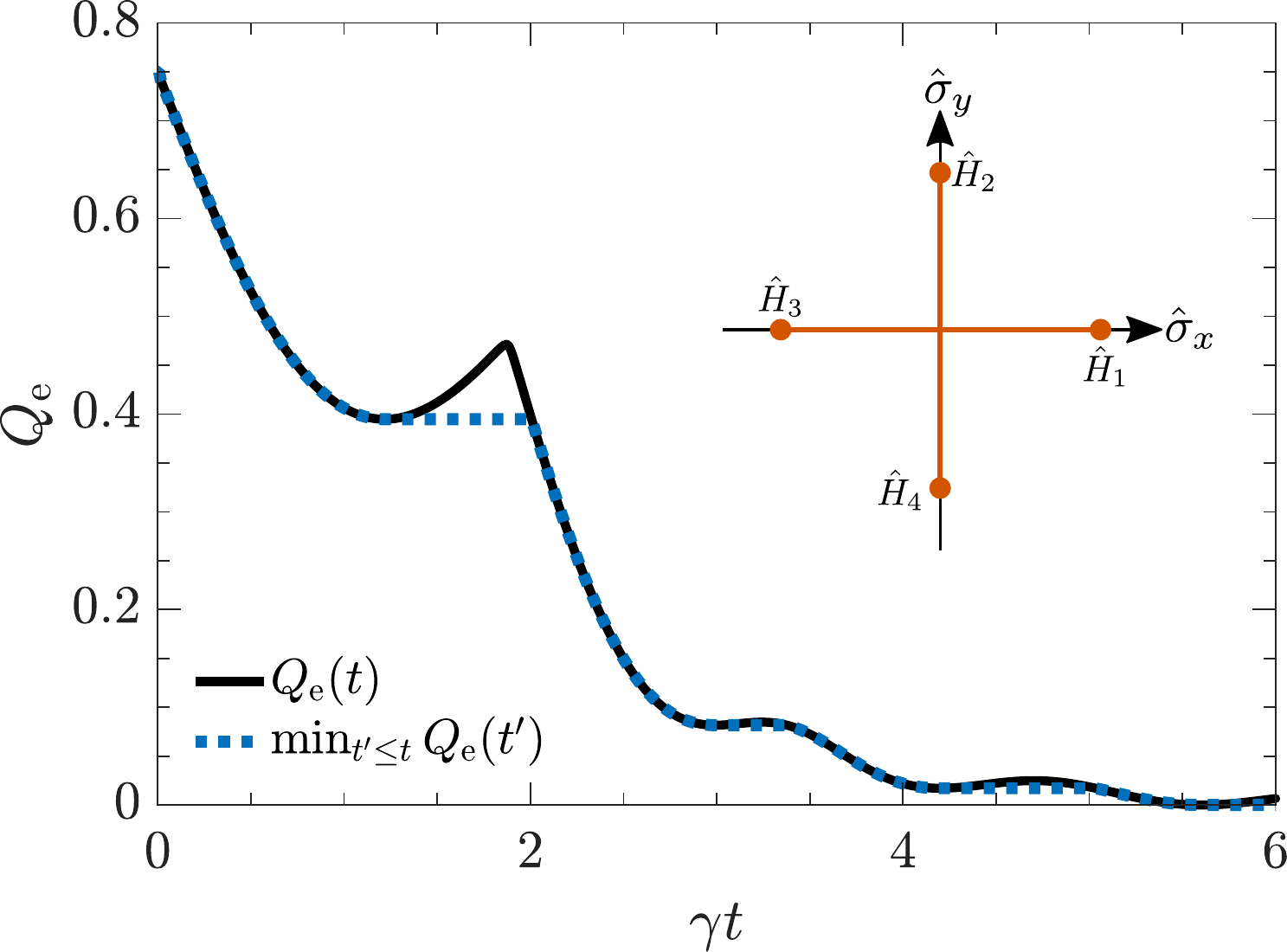}
\caption{The phase of a Rabi drive on a two-level system \eqref{eq:RabiDrive} is determined by performing a combined measurement on the atom and the emitted field.
Inset: The four hypotheses are $\H_1=\Omega\sx/2$, $\H_2=\Omega\sy/2$, $\H_3=-\Omega\sx/2$ and $\H_4=-\Omega\sy/2$.
Main figure: The probability of assigning a wrong hypothesis is shown for
$\Omega = 2\gamma$ and an atom initialized in its ground state. The dotted curve shows the smallest error probability
reachable by any given time $t$.
}
\label{fig:RabiExample}
\end{figure}

We first examine a two-level system driven resonantly with a known Rabi frequency $\Omega$ but with an unknown complex phase $\phi_m$. In a frame rotating with the resonance frequency, the candidate Hamiltonians can be written as,
\begin{align}\label{eq:RabiDrive}
\H_m = \Omega(\cos(\phi_m)\sx+\sin(\phi_m)\sy)/2.
\end{align}
In this example we consider the phases $\phi_m = \pi(m-1)/2$ as illustrated in the inset of \fref{fig:RabiExample}, and we assume that the atomic excitation decays into the environment at a known rate $\gamma$ such that $\hat{c} = \sqrt{\gamma}\sm$. The Rayleigh component of the emitted radiation is in phase with the driving field and homodyne detection should thus gradually reveal the value of $\phi_m$.
In contrast, photon counting tracks the intensity of the emitted radiation and hence maps the excitation of the system which is independent of $\phi_m$.

The full curve in \fref{fig:RabiExample} shows the minimum error probability $\Qe(t)$ calculated from our theory as a function of the duration $t$ of the experiment.
Initially $\Qe(t=0) = 75\%$, reflecting that each of the four hypotheses are a priori equally probable ($P_m = 1/4$) while for large times they may be unambiguously discriminated. Interestingly, however, $\Qe(t)$ is not a monotonous function of $t$. For instance, it reaches a minimum around $\gamma t \simeq 1.25$ and if for some reason the experiment lasts a little longer, $\gamma t \lesssim 2$, the ability to discriminate the four cases deteriorates. The reason is that, irrespective of the excitation phase, the atomic Bloch vector approaches the vertical direction around $\Omega t \simeq \pi$ , and only the emitted radiation provides any information until the Bloch vector candidates evolve further. It is thus sometimes favourable to perform a measurement on the atom and the field at an earlier time and keep the result rather than wait for more data to accumulate. The dotted curves tracks the minimum error probability obtainable by performing a measurement before any given time $t$ and hence represents the lowest error achievable in an experiment of duration $t$.

\subsection{Number of atoms inside a cavity}
\label{sec:Cavity}
Here we imagine a cavity field driven by a (classical) field of strength $u$ and interacting with an unknown but small number of atoms. Due to an out-coupling at a rate $\kappa$ from the cavity, the emitted radiation from the spins can be monitored and we further assume that the experimental setup allows direct measurements on the atomic ensemble.
The atoms are modelled as $N$ two-level systems of which $m$ are coupled linearly with strength $g$ to a single cavity mode and $N-m$ are uncoupled.
Assuming the bad cavity limit, the field may be adiabatically eliminated leading to an effective Hamiltonian and relaxation of the atoms.
The different hypotheses concerning the number of spins inside the cavity are hence characterized by $N$ sets of Hamiltonians and relaxation operators,
\begin{align}\label{eq:AtomsCavity}
\begin{split}
\hat{H}_m &= g\left(\alpha \hat{S}^{(m)}_++\alpha^* \hat{S}^{(m)}_-\right)
\\
\hat{c}_m &= \sqrt{\gamma_p}\hat{S}^{(m)}_-,
\end{split}
\end{align}
where
$\hat{S}^{(m)}_\pm = \sum_{i=1}^m \hat{\sigma}_\pm^{(i)}$, $\alpha = 2u/\kappa$ and $\gamma_p = 4g^2/\kappa$ is the Purcell-enhanced decay rate.
We assume that the number of atoms coupled to the cavity is Possion distributed with mean value $\mu = 1.5$. The probabilities prior to the experiment are hence $P_m\propto \mu^m/m!$.


\begin{figure}
\includegraphics[trim=0 0 0 0,width=0.9\columnwidth]{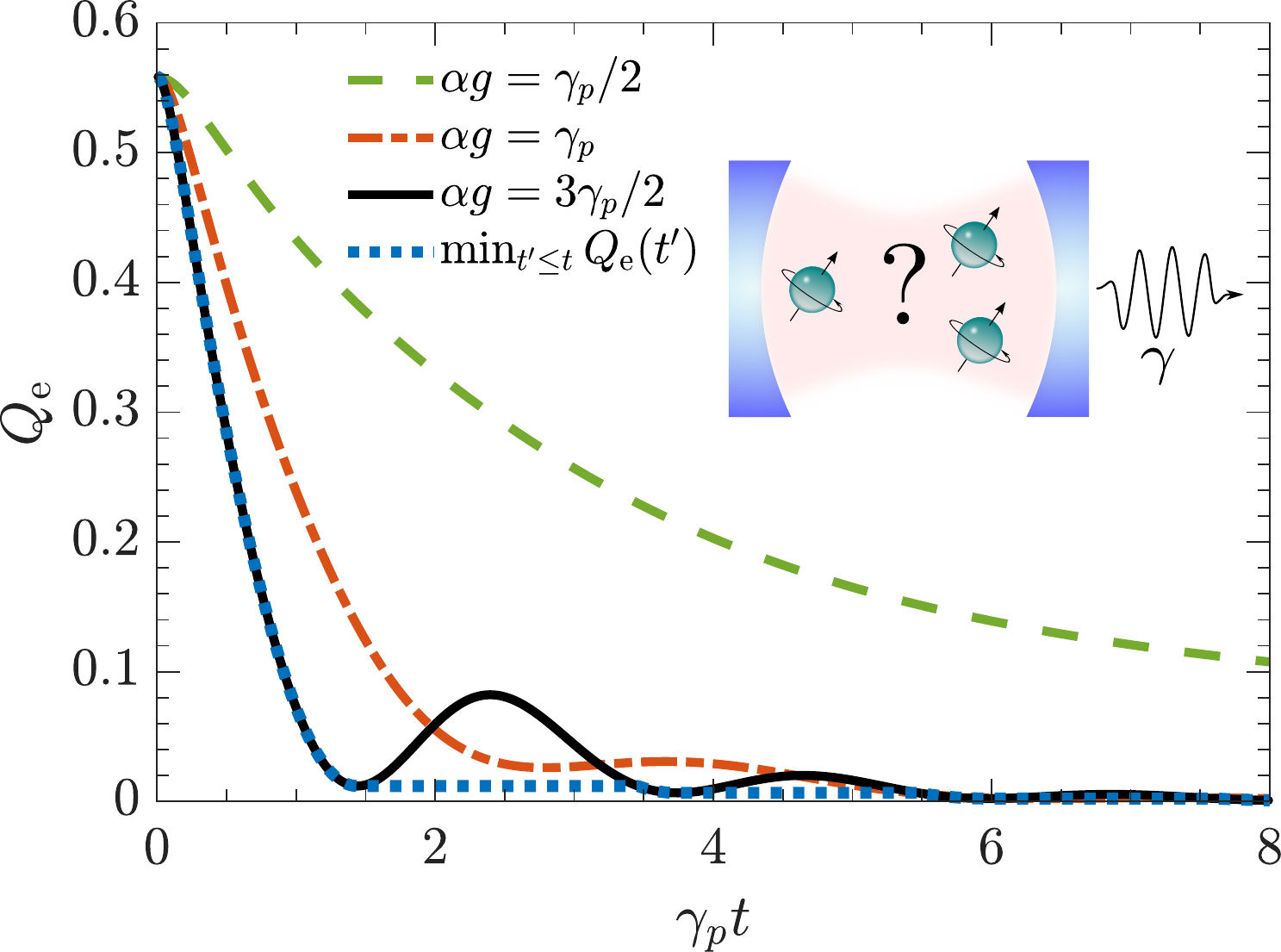}
\caption{Inset: An unknown number of atoms are coupled to a driven cavity. The number is estimated by performing a combined measurement on the atomic spins and the field emitted from the cavity. The hypotheses are $m=1,2,3$ and $4$ atoms and the system evolves according to the Hamiltonian (\ref{eq:AtomsCavity}). Main figure: The probability of assigning a false hypothesis is shown for weak driving $\alpha g = \gamma_p/2$, for moderate driving $\alpha g = \gamma_p$ and for strong driving $\alpha g = 3\gamma_p/2$. The dotted curve shows the smallest error probability
reached before any given time $t$ for the under damped cases where local minima occur in $\Qe(t)$.
}
\label{fig:QeAtomsInCavity}
\end{figure}

\fref{fig:QeAtomsInCavity} shows the evolution of the minimum error probability in distinguishing the cases of $m = 1,2,3$ and $4$ atoms inside the cavity.
Results are shown for a weakly, a moderately and a strongly driven cavity, respectively. While it is never favourable to drive the cavity very weakly since this does not lead to a florescence signal with much structure, it is evident that the moderate driving case outperforms the strong driving case for a brief period around $\gamma_p t \simeq 2.5$. However, the dotted curves, tracking the minimum error probability obtainable by performing a measurement at an optimal time before $t$, shows that the stronger driving is always favourable.

\subsection{Relative positions of a dopant ion}
\label{sec:Lattice}
\begin{figure*}
\includegraphics[trim=0 0 0 0,width=2\columnwidth]{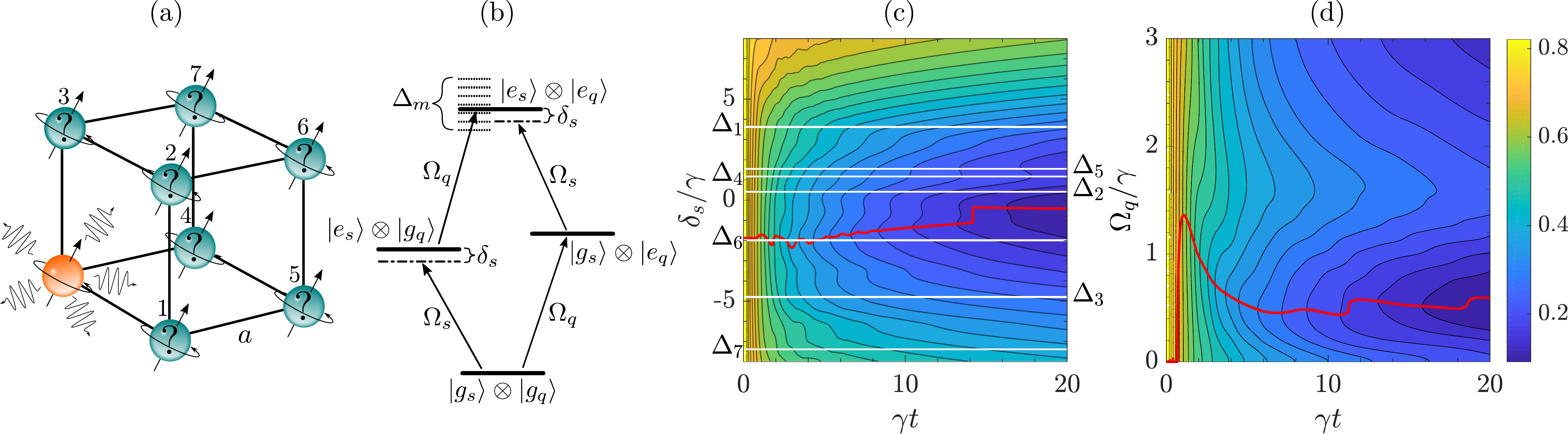}
\caption{
(a)
The position of a qubit ion (green spin) in a cubic lattice structure with lattice constant $a$ affects the fluorescence signal emitted by a sensor ion (orange spin) of a different species at a nearby lattice position and the internal state of the two ions.
In our example $\vec{\mu}_{s}=\vec{\mu}_{q} = (0.5,0.3,0.8)^T$ such that each of the seven possible marked positions yield a distinct energy shift of the sensor ion as given by \eqref{eq:DeltaOmega}.
(b)
The level structure of the Hamiltonian (\ref{eq:lattice}) of the two ions. The qubit ion is resonantly driven with a Rabi frequency $\Omega_q$ and the sensor ion is driven with a Rabi frequency $\Omega_s$ and a detuning $\delta_s$ from resonance. The sensor excited state is shifted by $\Delta_m$ due to the excited qubit ion.
(c)
Contour plot showing the probability $\Qe$ of assigning a wrong lattice position to the qubit ion as a function of time and the (constant) detuning $\delta_s$ of the sensor ion drive. The red line tracks the constant value of $\delta_s$ which minimizes $\Qe(t)$ at any given time. The white lines mark the energy shifts $\Delta_m/\gamma$ associated with each of the seven possible positions in (a).
(d)
Contour plot showing the probability $\Qe$ of assigning a wrong lattice position as a function of time and the (constant) strength of the qubit drive $\Omega_q$ with $\delta_s=0$. The red line tracks the constant value of $\Omega_q$ which minimizes $\Qe(t)$ for any given probing time.
Results in (c) and (d) are shown for $\Omega_s = 3\gamma$,
$\left(\frac{\epsilon+2}{3\epsilon}\right)^2\frac{ \mu_s \mu_q}{4\pi \epsilon_0} = 5\gamma a^3$ and equal priors $P_m = 1/7$.
}
\label{fig:lattice}
\end{figure*}
Our final example concerns testing of the relative positions between two impurity dipoles in a lattice structure. Rare-earth-ion dopants in inorganic crystals have permanent electric dipole moments which are different depending on whether each ion is excited or not and may hence be used for controlled gates in a quantum computation \cite{OHLSSON200271,PhysRevA.75.012304}. Such system are produced by low random doping during crystal growth, and for applications in a quantum sensor or computer, one may want to assess the relative positions of the individual ions (qubits) by probing their interactions.
A generic versions of this kind of setup in a cubic lattice structure with lattice constant $a$ is illustrated in \fref{fig:lattice}(a).
We assume dilute doping such that a read-out (sensor) ion couples only to a single qubit ion and we introduce a simple model of each ion as a two-level system. We let the sensor ion relax radiatively at a rate $\gamma$, and assume the qubit ion states to be long lived.
To obtain a florescence signal, the sensor is driven by a laser field with Rabi frequency $\Omega_s$ and detuned by $\delta_s$ from the bare transition frequency. We probe also the possibility to resonantly drive the qubit ion with a Rabi frequency $\Omega_q$ in order to optimize the sensing capabilities.
The candidate Hamiltonians may hence be written,
\begin{align} \label{eq:lattice}
\begin{split}
\hat{H}_m &= \frac{\Omega_s}{2}\sx^{(s)}+\frac{\Omega_q}{2}\sx^{(q)}-\delta_s\ket{e_s}\bra{e_s}
\\
&+\Delta_m \ket{e_q}\bra{e_q}\otimes \ket{e_s}\bra{e_s},
\end{split}
\end{align}
where the latter term accounts for the state dependent shift in frequency of the sensor ion due to the dipole-dipole interaction with the qubit ion \cite{jackson2007classical},
\begin{align}\label{eq:DeltaOmega}
\Delta_m = \left(\frac{\epsilon+2}{3\epsilon}\right)^2\frac{ \mu_s \mu_q}{4\pi \epsilon_0 r_m^3}\left[\hat{\vec{\mu}}_s\cdot\hat{\vec{\mu}}_q-3(\hat{\vec{\mu}}_s\cdot\hat{\vec{r}}_m)(\hat{\vec{\mu}}_q\cdot\hat{\vec{r}}_m)\right].
\end{align}
Here $\vec{\mu}_{s(q)} = \mu_{s(q)}\hat{\vec{\mu}}_{s(q)}$ is the difference in permanent electric dipole moment between the excited and ground state of the sensor (qubit) ion and $\vec{r}_m = r_m\hat{\vec{r}}_m$ is the vector between the sensor and the qubit with which the hypothesis testing is concerned.
The prefactor,
where $\epsilon$ is the relative permittivity at zero frequency, accounts for local field corrections due to the crystal host material.
One example is Eu$^{3+}$
or Pr$^{3+}$ ions doped in an YAlO$_3$ or an Y$_2$SiO$_5$ crystal \cite{PhysRevB.55.11225}. A suitable sensor ion could be Ce$^{3+}$
which has a large difference $\vec{\mu}_{s}$ in static dipole moment \cite{ahlefeldt2011characterisation}.
A level diagram for the Hamiltonian (\ref{eq:lattice}) is shown in \fref{fig:lattice}(b) and the level shift as well as the driven transitions are indicated.

Imagine first that we prepare the qubit ion in the excited state and then turn off the qubit drive ($\Omega_q=0$). As illustrated in \fref{fig:lattice}(b) the resonance frequency of the excited state is then shifted by $\Delta_m$ in a manner depending on the position of the qubit ion.
It is intuitively clear that a higher sensitivity can be obtained if the system is driven at the actual resonance. I.e. by detuning the sensor ion driving laser such that $\delta_s$ matches the true $\Delta_m$.
In \fref{fig:lattice}(c), the contour plot shows the error probability as a function of time and the (constant) value of $\delta_s$. The red curve tracks the optimum which is seen to be located near the mean value of the $\Delta_m$. Based on this understanding, one could imagine an optimized scheme where $\delta_s$ is cycled though the different $\Delta_m$ candidates with an appropriate portion of the total experimental time allocated to each.

By preparing an excited qubit the last term in \eqref{eq:lattice} remains fully active at all times while driving the qubit yields a transient evolution of the system which might depend more strongly on $\Delta_m$.
To investigate this, we show in \fref{fig:lattice}(d) a contour plot of the error probability as a function of time and the qubit drive strength. The red line tracks the (constant) values of $\Omega_q$ which minimizes $\Qe(t)$ at any given time.
Interestingly, for short times a relatively strong drive is favourable. This can be explained by the same mechanisms as in the previous example: When the system is strongly driven, it undergoes oscillations which are more pronounced at short times and whose frequency is modulated by $\Delta_m$. For longer times it becomes favourable to maintain the excited state for longer durations and with the parameters used here the optimal value lies around $\Omega_q \simeq 0.5\gamma$.

We presented this example for dopant ions but the general idea may be relevant in a number of similar setups. For example, the dipole-dipole potential between neutral atoms similarly yields an energy shift of the form \eqref{eq:DeltaOmega} which is responsible for, e.g., the Rydberg Blockade mechanism \cite{PhysRevLett.85.2208}. Hence, our formalism could  be fitted to the determination of the relative positions of Rydberg atoms in an optical lattice structure.
Another platform is the sensing of a remote nuclear spin $\hat{I}$ by an electron spin $\hat{S}$. Here the frequency shift is due to the hyperfine coupling
\cite{childress2006coherent,PhysRevB.78.094303,PhysRevLett.109.137602}.
For instance the electron spin of an NV centre can be used to sense the position of a $^{13}$C impurity in a vapour deposited (CVD) diamond which has a $^{13}$C abundance of less than $0.01\%$ \cite{zhao2012sensing}.

\section{Conclusion and outlook}
\label{sec:5}
We have outlined how optimal discrimination between an arbitrary number of quantum states may be phrased as a semi-definite programming problem for which efficient numerical solutions are available.
Our example, considering three non-orthogonal state of a two level system, suggests that more generally, 
the probability of assigning a false state is minimized by already disregarding a subset of the candidate states,
based on their overlaps and preparation probabilities,
prior to the discriminating measurement.

We then utilize that distinguishing a set of $N$ hypotheses for the evolution of a Markovian open quantum system is equivalent to the discrimination of a set of time dependent states of the full system and its environment. We show how their overlaps may be calculated in a straightforward manner and used to construct a lower dimensional representation which is suitable for numerical treatment. 
This allows us to evaluate a lower (quantum) bound to the probability of assigning a false hypothesis, and the three examples presented in this article serves to illustrate some insights that may be obtained by such an analysis.

For the example in Section~\ref{sec:Lattice}, we show how different but constant values of the qubit ion Rabi frequency and the detuning in the driving frequency of the sensor ion, lead to different error probabilities. In optimal control, the bound could be further optimized by allowing time dependent parameters $\Omega_q(t)$ and $\delta_s(t)$.
More generally, by following this line of thought our formalism is suitable for systematic optimization of the sensing capabilities in a given quantum setup by e.g. controlling axillary Hamiltonian parameters or environmental coupling strengths.    

Finally, we want to emphasize that our quantum bound may pertain to a highly non-local measurement performed on the full state of the system and its environment. Such a measurement is in general infeasible to implement and in real experimental situations one has recourse to perform a more conventional measurement of the environment. For instance, the radiation emitted by an open system may be monitored by photon counting or homodyne demodulation \cite{PhysRevA.89.052110,PhysRevA.87.032115,PhysRevA.95.022306,PhysRevA.94.032103,PhysRevA.95.022110}, and the signal possibly combined with a final projective measurement on the system.
Depending on the setup, different monitoring schemes are favourable as discussed in relation to the example of Section~\ref{sec:Rabi}.
See e.g. Refs.~\cite{PhysRevLett.108.170502,PhysRevA.91.012119,inprep} for an investigation of hypothesis testing with continuous measurements.

\section{Acknowledgements}
The authors acknowledge financial support from the Villum Foundation and the European Union’s Horizon 2020 research and
innovation programme under grant agreement No 712721 (NanOQTech).
A. H. K. further acknowledges financial support from the Danish Ministry of Higher Education and Science.


%

\end{document}